\documentclass[aps,prl,twocolumn,superscriptaddress,unsortedaddress]{revtex4}
\usepackage{epsfig}
\usepackage{amsmath}

\newcommand{\be}{\begin{equation}}
\newcommand{\ee}{\end{equation}}
\newcommand{\bk}{{{\bf{k}}}}

\newcommand{\bea}{\begin{eqnarray}}
\newcommand{\eea}{\end{eqnarray}}
\newcommand{\ra}{\rangle}
\newcommand{\la}{\langle}

\newcommand{\bS}{{\bf S}}

\newcommand{\dg}{{\dagger}}
\newcommand{\pdg}{{\phantom\dagger}}

\newcommand{\hb}{\hat{b}}
\newcommand{\ha}{\hat{a}}

\begin{document}

\title{Quantum paramagnets on the honeycomb lattice and field-induced
N\'eel order: Possible
application to Bi$_3$Mn$_4$O$_{12}$(NO$_3$)}
\author{R. Ganesh}
\affiliation{Department of Physics, University of Toronto, Toronto, Ontario M5S 1A7, Canada}
\author{D. N. Sheng}
\affiliation{Department of Physics and Astronomy, California State University, Northridge, California 91330, USA}
\author{Young-June Kim}
\affiliation{Department of Physics, University of Toronto, Toronto, Ontario M5S 1A7, Canada}
\author{A. Paramekanti}
\affiliation{Department of Physics, University of Toronto, Toronto, Ontario M5S 1A7, Canada}
\affiliation{Canadian Institute for Advanced Research, Toronto, Ontario, M5G 1Z8, Canada}

\date{\today}

\begin{abstract}
{
%Recent experiments show that Bi$_3$Mn$_4$O$_{12}$(NO$_3$), a
%spin-3/2 bilayer honeycomb lattice antiferromagnet,
%remains paramagnetic to the lowest temperature in zero magnetic field but develops
%long-range N\'eel
%order beyond a critical field.  
Motivated by recent experiments on the spin-3/2 frustrated
bilayer honeycomb antiferromagnet Bi$_3$Mn$_4$O$_{12}$(NO$_3$),
we study the spin-S Heisenberg model on the honeycomb lattice with various additional
exchange interactions
which frustrate N\'eel order. Using spin wave theory, exact diagonalization, 
and bond operator theory, we consider the effects of (i) second-neighbor exchange, (ii)
biquadratic exchange for $S = 3/2$ which
leads to an AKLT valence bond solid, and (iii)
bilayer coupling which leads to an interlayer dimer solid. We show that the resulting
paramagnetic states undergo a transition to N\'eel order beyond a critical magnetic field.
We discuss experimental implications for Bi$_3$Mn$_4$O$_{12}$(NO$_3$).}
%and discuss disorder and spin-orbit effects.}
\end{abstract}

\maketitle

The 
interplay of quantum mechanics and frustrated interactions 
in quantum magnets leads to a variety of remarkable 
phases including spin liquid Mott insulators, valence bond crystals, and Bose-Einstein condensates
of magnons
\cite{reviews}.
Experiments on Bi$_3$Mn$_4$O$_{12}$(NO$_3$) 
indicate that this material is a possible new candidate for a quantum spin liquid \cite{BiMnO}.
The octahedral crystal field, together with strong Hund's coupling, leads
to Heisenberg-like spin-3/2 moments on the
Mn$^{4+}$ ions which form a bilayer honeycomb lattice. Despite the bipartite structure, and a large
antiferromagnetic Curie-Weiss constant $\Theta_{CW}\approx -257K$, this system shows no
magnetic order (or any other phase transition)
down to $T\!\sim\! 1 K$ \cite{BiMnO}. 
This observation hints at
frustrating interactions which may
lead to interesting paramagnetic ground states \cite{fouet2001,mattsson1994,takano2006,assaad,mulder2010,kawamura2010,fawang2010,jafari2010}.

Recent neutron scattering experiments \cite{unpub} on powder samples  of Bi$_3$Mn$_4$O$_{12}$(NO$_3$) 
in zero magnetic field indicate that there are short range spin correlations in this material,
with some antiferromagnetic coupling between the two layers forming the bilayer, but negligible
interactions
between adjacent bilayers. Remarkably, a critical magnetic field, $B_c \sim 6 $ Tesla, leads to sharp
Bragg spots consistent with
three dimensional (3D) N\'eel order \cite{unpub}. 
Motivated by the observation that
the field required to induce N\'eel order appears to extrapolate to a nonzero value at
$T=0$, we propose that this system could exhibit a field tuned quantum phase transition
into the N\'eel state. We flesh out this idea by studying various interactions which
could frustrate the N\'eel order in this material.

We first examine the possibility that the N\'eel order in Bi$_3$Mn$_4$O$_{12}$(NO$_3$) is destroyed
by a
frustrating second-neighbor exchange ($J_2$) in addition to the dominant nearest neighbor
term ($J_1$). Sidestepping the issue of
what state results from such quantum melting,
we study the magnetic field dependence of the critical $J_2/J_1$ required to destroy the N\'eel order.
Using spin-wave theory, we show that a nonzero magnetic field enhances the critical $J_2/J_1$,
%, this
%effect being more pronounced for small spin values.
opening up a regime where applying a critical
field to the non-N\'eel state yields long-range N\'eel order.

We next explore other frustrating interactions which might kill N\'eel order and lead to novel quantum paramagnetic ground states on the honeycomb lattice. 
We focus here on two valence bond solid (VBS) states, which do not break any symmetries and are expected to show no
thermal phase transitions as is the case with Bi$_3$Mn$_4$O$_{12}$(NO$_3$).
(i) Motivated by Bi$_3$Mn$_4$O$_{12}$(NO$_3$), we study
a generalized spin-3/2 model including biquadratic and bicubic spin interactions which
permits an Affleck-Kennedy-Lieb-Tasaki (AKLT)
ground state \cite{AKLT1987,arovas1988,KLT1988}.
Using exact diagonalization (ED) to compute the fidelity susceptibility \cite{alet2010},
we show that this model exhibits a direct N\'eel-AKLT transition.
We also obtain the
spin gap at the AKLT point. (ii) In view of the fact that Bi$_3$Mn$_4$O$_{12}$(NO$_3$) consists of stacked
bilayers, we use a spin-S generalization \cite{brijesh} of the bond operator formalism
\cite{sachdevbhatt1990}
 to show that a sufficiently strong 
bilayer coupling leads to an interlayer VBS state.
Both valence bond solids, the interlayer VBS and the AKLT state, are shown to
undergo a magnetic field induced quantum phase transition into a state which
exhibits
N\'eel order.
Our results on the AKLT state are of broader interest given
recent proposals to use this
state as a universal quantum computation resource \cite{tcwei}.

Finally, we discuss possible
experiments on Bi$_3$Mn$_4$O$_{12}$(NO$_3$)
which may help to discriminate between the various states 
we have studied, and to distinguish them from possible Z$_2$ spin liquids
\cite{fawang2010}.

{\it Second-neighbor exchange.---}
It has been suggested that the absence of N\'eel order in Bi$_3$Mn$_4$O$_{12}$(NO$_3$) is linked to
non-negligible further neighbor interactions \cite{BiMnO}. We therefore study a minimal Hamiltonian,
\bea
\label{Eq:Hnnexch}
H = J_1 \sum_{\la i j\ra} \bS_i \cdot \bS_j + J_2  \sum_{\la\la i j \ra\ra} \bS_i \cdot \bS_j  - B
\sum_i S_{i}^{z}
\eea
where $\la.\ra$ and $\la\la.\ra\ra$ denote nearest and next-nearest neighbor 
bonds respectively, and $B$ is a
Zeeman field. Let us begin with a classical analysis valid for $S\!=\!\infty$.
When $J_2\!\!=\!\!B\!\!=\!\!0$,
the ground state has collinear N\'eel order.
For $J_2\!\!=\!\!0$ and $B\!\!\neq \!\!0$,
the spins in the N\'eel state start off in the plane perpendicular to the
applied field and cant along the field direction until they are
fully polarized for $B \!> \!6 J_1 S$.
For $B \!<\! 6 J_1 S$, the spin components transverse to the magnetic field
have staggered N\'eel order for $J_2\! < \! J_1/6$; this gives way to a one-parameter family of
degenerate (canted) spirals for $J_2\!>\!J_1/6$\cite{mulder2010}.

\begin{figure}[tb]
\includegraphics[width=3.3in]{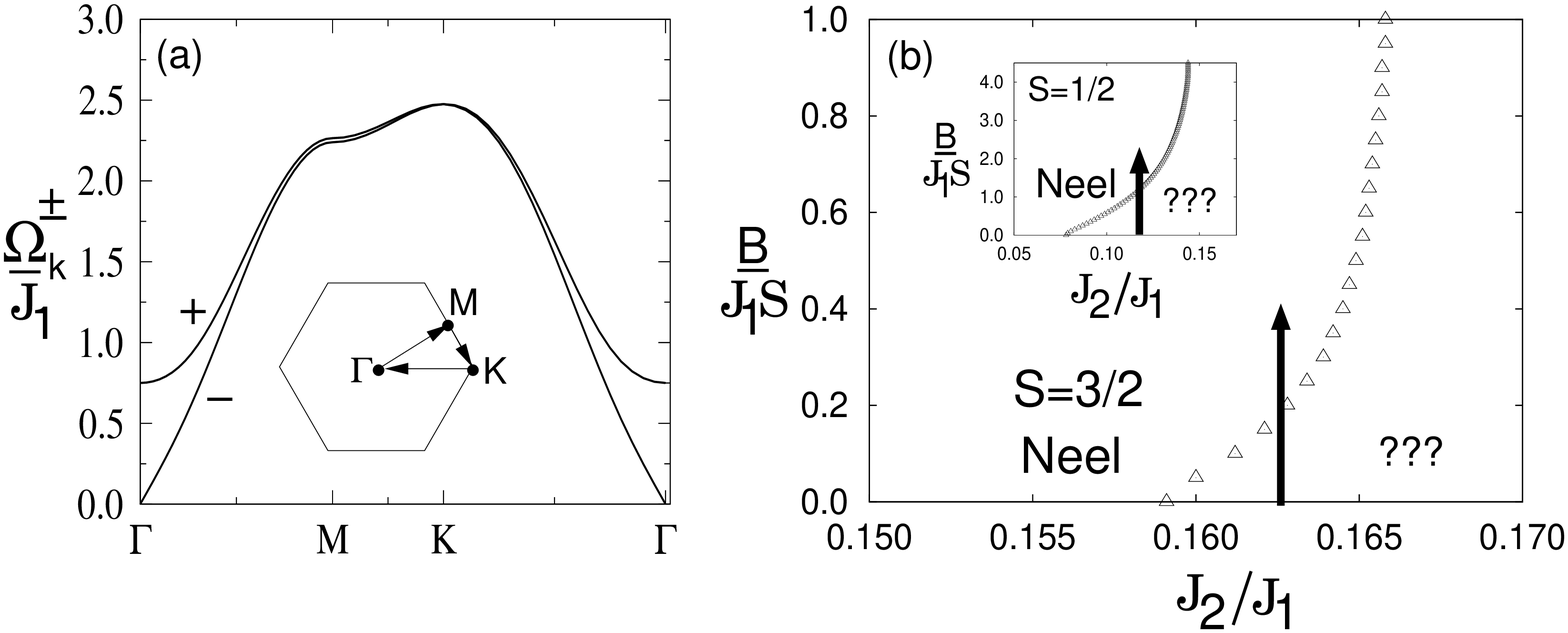}
\caption{(a) Dispersion of magnon modes $\Omega_{\bk}^\pm$ in the $J_1$-$J_2$ model along 
depicted path in the Brillouin zone 
for $J_2\!\!=\!\!0.15 J_1$, $S\!\!=\!\!3/2$ and $B\!\!=\!\!0.5 J_1 S$. 
(b) $T\!\!=\!\!0$ melting of N\'eel order for $S\!\!\!=\!\!\!3/2$ in the 
$B$-$J_2$ plane
(open triangles) obtained using a Lindemann-like criterion, $\sqrt{\la S_x^2\ra \!-\! \la S_x\ra^2} \!\! = \!\!  3 \la S_x\ra$. 
The region ``???'' is a quantum disordered state -
possibly a valence bond solid or a quantum spin liquid.
Arrow depicts path along which one obtains
a field-induced transition to N\'eel order. Inset depicts a similar melting curve for $S\!=\!1/2$.
}
\label{fig:phaseboundary}
\vskip -0.1in
\end{figure}

Incorporating quantum fluctuations is likely to lead to melting of N\'eel order even for $J_2 \! < \! J_1/6$.
Such fluctuations are also likely to completely suppress the classical spiral order \cite{mulder2010}.
%leading to interesting quantum paramagnets.
Using spin wave theory, we argue here that a small nonzero $B$ enhances the stability of the
N\'eel order compared to the zero field case. (i) For small nonzero $B$,
spin canting leads to a small decrease, $\propto B^2$, in the classical
staggered magnetization transverse to the field.
(ii) On the other hand,
one of the two magnon modes (labelled $\Omega_\bk^{+}$) 
acquires a nonzero gap $\propto B$ at the $\Gamma$-point
as shown in Fig.~\ref{fig:phaseboundary}(a)
(for $S\!=\!3/2$ with $J_2\!=\!0.15 J_1$ and $B\!=\!0.5 J_1 S$).  This
suppresses quantum fluctuations.  For $B \!\!\ll\!\! 6 J_1 S$, the latter effect
overwhelms the former, leading to enhanced stability of N\'eel order. 

To estimate the `melting curve', we assume that the transverse spin components have N\'eel order
along the $S_x$-direction, and 
use a heuristic Lindemann-like criterion for melting:
$\sqrt{\la S_x^2\ra - \la S_x\ra^2} \! > \! \alpha \la S_x \ra$ where the expectation values are evaluated
in linear spin wave theory (see Supplementary Information for
details). As shown in Fig.\ref{fig:phaseboundary}(b) and its inset, quantum fluctuations at $B=0$ 
lead to melting of N\'eel order even for $J_2 \!<\! J_1/6$ (i.e., before the classical destruction of 
N\'eel order). We set $\alpha\!\!=\!\!3$ since this leads to a
melting of N\'eel order for $S\!\!=\!\!1/2$ at $J_2 \approx 0.08 J_1$, in agreement
with a recent variational Monte Carlo study by Clark {\it et al} \cite{fawang2010}.

For nonzero $B$, the `melting point'
moves towards larger $J_2$, leading to a window of $J_2$
over which the quantum disordered liquid can undergo a field-induced
phase transition to N\'eel order (see Fig.~\ref{fig:phaseboundary}(b)).  This is
consistent with recent neutron diffraction experiments \cite{unpub} on Bi$_3$Mn$_4$O$_{12}$(NO$_3$).
The window of $J_2$ where such physics is operative appears to be small for $S\!\!=\!\!3/2$; however, 
disorder effects would suppress the stiffness \cite{disorderboson} and 
may enhance this regime. We expect field induced N\'eel order even for $S\!\!=\!\!1/2$ (see inset to Fig.~\ref{fig:phaseboundary}(b)). 
 This can be verified by including a magnetic field in calculations reported in Ref.\cite{fawang2010}.
Our results also explain recent Monte Carlo simulations
of the classical model with $B\!\!\neq\!\! 0$ \cite{kawamura2010} ; if $J_2\!\! = \!\!0.175 J_1$, as in the simulations,
a nonzero $B$ takes us closer to the melting curve, and may lead to the numerically
observed enhanced N\'eel correlations. Nevertheless, we expect that there will be no field-induced 
{\it long-range} N\'eel order for $J_2\!\! = \!\!0.175 J_1$ in the classical model.

Next-neighbor exchange thus provides a plausible explanation for the experimental
data on Bi$_3$Mn$_4$O$_{12}$(NO$_3$). We next turn to an exploration of other
mechanisms which frustrate N\'eel order.

{\it AKLT valence bond solid.---} A particularly
interesting spin-gapped ground state of a magnet with spin-$S$ atoms on a lattice of
coordination number $z\!\!=\!\!2S$, is an AKLT valence bond state.
Each spin-$S$ is viewed as being composed of $2S$ spin-1/2
moments symmetrized on-site, with each spin-1/2 moment forming a singlet with
one neighbor \cite{AKLT1987,arovas1988,KLT1988}. It was originally proposed as a realization
of Haldane's prediction of a spin-gapped ground state in 1D integer spin systems \cite{Haldane}.
Assuming that the Mn$^{4+}$ 
ions 
in Bi$_3$Mn$_4$O$_{12}$(NO$_3$) 
mainly interact with the three neighboring spins in the
same plane, this condition is satisfied with $S=3/2$ and $z=3$. The honeycomb
lattice AKLT state has
exponentially decaying spin correlations \cite{arovas1988}, and
it is the exact, and unique, zero energy ground state of the parent Hamiltonian
$
H_{\rm AKLT} = \sum_{\la ij\ra} P^{(3)}_{i,j}.
$
Here $P^{(\ell)}_{i,j}$ denotes a projector on to total spin-$\ell$ for a pair of  spins
on nearest neighbor sites $(i,j)$.
Denoting $T_{i,j}\!\equiv\! \bS_i \cdot \bS_j$, we find
$P^{(3)}_{i,j}= \frac{11}{128} +
\frac{243}{1440} T_{i,j}
+ \frac{116}{1440} T^2_{i,j}
+ \frac{16}{1440} T^3_{i,j}$.
We do not have a microscopic basis, at this point, for
such higher order exchange terms in Bi$_3$Mn$_4$O$_{12}$(NO$_3$); but
it is encouraging to note that the coefficients of such terms are smaller than the
leading Heisenberg interaction.

\begin{figure}[tbp]
\includegraphics[width=3.2in]{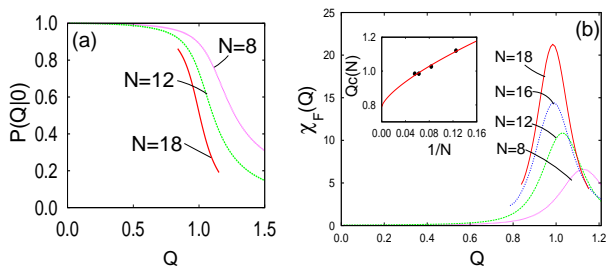}
\caption{(Color online) 
(a) Overlap $P(Q|0)$ of the ground state at $Q$ with the N\'eel state ($Q\!\!=\!\!0$) 
for various system sizes $N$, showing its rapid drop around the N\'eel-AKLT
transition. (b) Fidelity susceptibility $\chi_F(Q)$ versus $Q$ for various system sizes $N$, with
the peak indicating the N\'eel-AKLT transition point $Q_c(N)$.  Inset:
$Q_c(N)$ versus $1/N$, together with a fit $Q_c(N)\!=\!Q^\infty_c\!+\! b N^{-\frac{1}{2\nu}}$
(with a choice $\nu \approx 0.7$ assuming an $O(3)$ quantum phase transition in 2D)
which leads to $Q^\infty_c \approx 0.8$.}
\label{fig:fidelity}
\vskip -0.2in
\end{figure}

We have investigated, using ED on system sizes $N\!=\!12$-$18$, the phase diagram of a generalized spin-3/2
model,
\be
H_Q = 
\!=\! (1-Q) \sum_{\la ij\ra} 
\bS_i\!\cdot\! \bS_j
\!+\! g Q H_{\rm AKLT},
\label{HQ}
\ee
which interpolates between a Heisenberg model (at $Q\!\!=\!\!0$) and $gH_{AKLT}$ (at $Q\!\!=\!\!1$). 
We set $g\!\!=\!\!1440/243$, so that the coefficient of 
$\bS_{i}\!\cdot\!\bS_{j}$ is unity.
For $Q\!=\!0$, our analysis of the finite size spectrum shows
that the ground state energy $E_g(N,S^{\rm tot})$, as a function of
total spin $S^{\rm tot}$,
varies as $S^{\rm tot}(S^{\rm tot}\!+\!1)$,  in agreement with the expected Anderson tower
for a N\'eel ordered state. It is consistent with earlier work showing 
N\'eel order even for spin-1/2 \cite{fouet2001,young1989,baskaran2009}.
To establish the N\'eel-AKLT transition as a function of $Q$, we
study overlaps $P(Q|Q')\!\! =\!\! |\la \Psi_g(Q) | \Psi_g(Q') \ra|$ of the ground state
wave functions at $Q$ and $Q'$.
As shown in Fig.\ref{fig:fidelity}(a),
the overlap $P(Q|0)$, of the
ground state wavefunction at $Q$ with the N\'eel state at $Q'\!\!=\!\!0$, is nearly unity for
$Q \!\!\lesssim\!\! 0.8$, suggesting that the ground
state in this
regime has N\'eel character.
For $0.8 \!\! \lesssim \!\! Q \!\!< \!\!1.2$,
we observe a dramatic drop of $P(Q|0)$ for all system
sizes, which indicates a N\'eel-AKLT quantum phase
transition. To locate the
transition more precisely,
we compute the `fidelity susceptibility'
 $\chi_F(Q) \! = 2 (\!1\!-P(Q|Q+\delta))/\delta^2$, with $\delta \! \to\! 0$,
 which measures the 
change of the wavefunction when $Q\! \rightarrow\! Q\!+\!\delta$ \cite{alet2010}.
Fig.\ref{fig:fidelity}(b) shows a plot of $\chi_F(Q)$ (with $\delta\!=\!0.005$). 
We observe a peak in $\chi_F(Q)$ which indicates a phase transition; this
peak shifts and grows sharper with increasing $N$. Assuming the thermodynamic
transition is at $Q^\infty_c$, and that the peak position $Q_c(N)$
satisfies the scaling relation 
$(Q_c(N)\!-\!Q^\infty_c) \!\!\sim \!\!N^{-1/2\nu}$, with $\nu \!\approx\!0.7$
for an $O(3)$ quantum phase transition\cite{square} corresponding to triplon condensation,
we estimate
the transition point
$Q^\infty_c \!\! \approx \! 0.8$.

\begin{figure}[tb]
\includegraphics[width=3.4in]{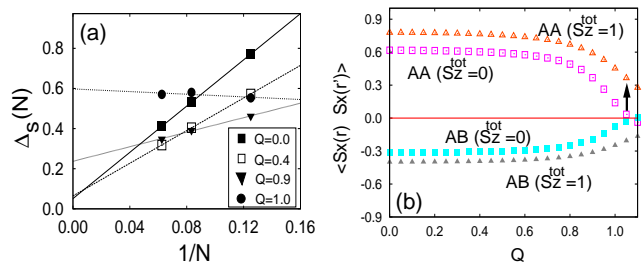}
\caption{(Color online) 
(a) Spin gap, $\Delta_s(N)$, versus $1/N$ for various $Q$,
with fits to the form $\Delta_s(N)=\Delta_s^\infty+b/N$. The small values of $\Delta_s^\infty$
for $Q\!=\!0.0,0.4$ are consistent with a gapless N\'eel state. For $Q\!=\!0.9,1.0$, the data
are consistent with a 
robust spin gap $\Delta_s^\infty$.
(b) $S_x$-spin correlations between distant sites on the same ($AA$) and
opposite ($AB$) sublattices
for $N\!\!=\!\!16$ system. The spin correlation is Neel-like ($\pm$) for all $Q$ shown; 
in the spin gapped AKLT state at Q$\sim$1, it short ranged and weak in the $S^{\rm tot}_z\!=\!0$ ground state 
but it is strongly enhanced (see arrow) in the $S^{\rm tot}_z\!=\!1$ case.}
\label{fig:corr}
\vskip -0.1in
\end{figure}

The spin gap $\Delta_s(N)
\!=\! E_g(N,S^{\rm tot}\!=\!1)\!-\!E_g(N,S^{\rm tot}\!=\!0)$ is plotted in
Fig.\ref{fig:corr}(a) for various $Q$ as a function
of $1/N$. Assuming a finite size scaling form
$\Delta_s(N) = \Delta_s^\infty + b/N$, we find a small value for $\Delta_s^\infty$ 
for $Q=0.0,0.4$, consistent with a gapless N\'eel state, while for $Q=0.9,1.0$
there appears to be a robust spin gap as $1/N \to 0$.
At the AKLT
point ($Q\!\!=\!\!1$),  we estimate $\Delta^\infty_s \approx 0.6$.

Since the spin gap is finite for $Q\! >\! Q_c$, we expect that applying a critical field $B_{c}
\! \propto\! \Delta_s$ will lead to a phase transition; the correlation functions of
the $S_z^{\rm tot}\!\!=\!\!1$ state at zero field will then reflect the correlations of the ground state 
for $B_z \!\!>\!\! B_{c}$. We plot, in Fig.\ref{fig:corr}(b), the
spin correlations on two maximally separated sites (for $N\!=\!16$) as a function of $Q$,
and make the following observations.
(i) For $S_z^{\rm tot}=0$, the ground state also has $S^{\rm tot}=0$, and
$\la S_x(i)S_x(j)\ra\!=\!\la S_z(i)S_z(j)\ra$ due to spin rotational 
invariance. At long distance, the spin correlation is strong in the N\'eel phase,
but drops rapidly to small values upon entering the AKLT state.
(ii)  In the $S_z^{\rm tot}=1$ sector,
$\la S^z(i)S^z(j) \ra \neq \la S_x(i)S_x(j) \ra$.
Remarkably, in this sector, as opposed to $S^{\rm tot}\!\!=\!\!0$,
we find a strong enhancement of {\it only} transverse correlations $\la S_x(i)S_x(j) \ra$ 
between
distant sites in the
AKLT state;
this finite-size result  suggests that the AKLT state will undergo, beyond a critical field,
a transition into a state with
in-plane N\'eel order.

{\it Interlayer valence bond solid. ---} The Mn sites in a unit cell of Bi$_3$Mn$_4$O$_{12}$(NO$_3$) form an AA
stacked bilayer honeycomb lattice. If the interplane antiferromagnetic
exchange $J_c$ is strong compared to the in-plane nearest neighbor exchange $J_1$, adjacent
spins on the two layers could dimerize and lead to loss of N\'eel order.
To study this interlayer VBS, we begin from the limit $J_1\!\!=\!\!0$; this leads
to the spectrum $E_j \! =\! -J_c (S (S+1)\! -\! j (j \!+\! 1)/2)$, with $j\!=\!0,1,\ldots,2S$ denoting the total spin state
of the dimer.
Restricting attention to the low energy Hilbert
space spanned by the singlet and the triplet states, we define generalized
spin-S bond operators via: $|s\ra = s^\dagger|0\ra$, and $|\alpha\ra=t^\dagger_\alpha|0\ra$, where $|0\ra$ is
the vacuum, and $|\alpha(\!=\!x,y,z)\ra$
are related to the $m_j$ levels of the triplet by $|z\ra\!=\!|m_j\!=\!0\ra$, 
$|x\ra \!=\! (|m_j\!=\!-1\ra\!-\!|m_j\!=\!1\ra)/\sqrt{2}$,
and $|y\ra \!=\! i (|m_j\!=\!-1\ra\!+\!|m_j\!=\!1\ra)/\sqrt{2}$. Denoting the two spins constituting the dimer, by $\bS_\ell$,
with layer index $\ell=0/1$,
we obtain \cite{brijesh}
\bea
\bS^\alpha_{\ell} \!\!&\!\approx\!&\!\! (-1)^\ell \sqrt{\frac{S(S+1)}{3}} (s^\dagger t_\alpha^\pdg \!+\! t^\dagger_\alpha s^\pdg) \!-\! 
\frac{i}{2} \varepsilon_{\alpha\beta\gamma} t^\dagger_\beta t^\pdg_\gamma,
\eea
together with the constraint $s^\dagger s^\pdg \!+t^\dg_\alpha t^\pdg_\alpha \!=\! 1$
at each site.

\begin{figure}[tb]
\includegraphics[width=2.2in]{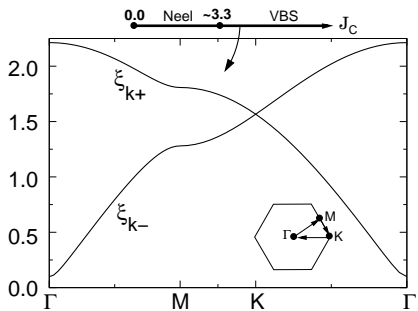}
\caption{Phase diagram of the $S\!=\!3/2$ bilayer honeycomb model obtained using
bond operator theory, and 
triplon dispersion along depicted path in the Brillouin 
zone within the interlayer
VBS state for $J_c/J_1 \!=\!3.8$ (in units where $J_1\!=\!1$).}
\label{fig:triplon}
\vskip -0.1in
\end{figure}

To treat the effect of $J_1$, we use bond operator mean field theory \cite{sachdevbhatt1990} which
yields a reasonably accurate phase diagram for the
spin-1/2 bilayer square lattice Heisenberg model \cite{square}.
Assuming the singlets are condensed in the dimer solid,
we replace $s^\dagger\!=\!s^\pdg\!=\!\bar{s}$, and incorporate a Lagrange multiplier in the Hamiltonian
which enforces $\la t^\dg_\alpha t^\pdg_\alpha \ra = 1-\bar{s}^2$ on average. Let $N$ be
the number of spins in each honeycomb layer.
We then obtain the Hamiltonian
$ H = \sum_{\alpha,\bk>0} \Psi^\dagger_{\bk\alpha}  M^\pdg_\bk \Psi^\pdg_{\bk\alpha} + 2 N C $, describing
the dynamics of the triplets. Here 
$\Psi^\dagger_{\bk\alpha}=(t^\dagger_{\bk \alpha 1} t^\dagger_{\bk \alpha 2} t^\pdg_{-\bk\alpha 1}
t^\pdg_{-\bk\alpha 2})$ (with $1,2$ denoting the two sublattices in each layer) and the 
matrix $M_\bk$ takes the form
\be
M_\bk = \begin{pmatrix} A_\bk & B_\bk & 0 & B_\bk  \\ B^*_\bk & A_\bk & B^*_\bk & 0  \\ 
0 & B_\bk & A_\bk & B_\bk \\
B^*_\bk & 0 & B^*_\bk & A_\bk \end{pmatrix}
\label{eq:Mk},
\ee
with
$A_\bk = J_c - \mu - J_c S (S+1)$ and $B_\bk = \frac{1}{3} \gamma_\bk J_1 S (S+1) \bar{s}^2$.
Here we have defined
$\gamma_\bk = 1 + {\rm e}^{-i \bk\cdot\hb} + {\rm e}^{-i \bk\cdot(\ha+\hb)}$,
with unit vectors $\ha=\hat{x},\hb=-\hat{x}/2+\sqrt{3} \hat{y}/2$, and the constant
$C= - \frac{\mu}{2} (\bar{s}^2-1) -  \frac{3}{4} (J_c - \mu - J_c S (S+1)) -  \frac{1}{2} J_c
\bar{s}^2 S(S+1)$.
Diagonalizing this Hamiltonian leads to the 
ground state energy per spin
$ E_g \!=\! \frac{3}{2 N} \sum_{\bk > 0} (\xi_{\bk +}  + \xi_{\bk -}) \!+\! C$
where $\xi_{\bk\pm} \!=\! \sqrt{ A_\bk (A_\bk \pm 2 |B_\bk|)}$.
Setting $\partial E_g/\partial \bar{s}^2\!=\!\partial E_g/\partial\mu\!=\!0$, we obtain the 
mean field values of $\bar{s}$ and $\mu$ which minimize the ground state energy
subject to the constraint. Solving these equations numerically, we find that the spin-$S$ interlayer
VBS is a stable phase for $J_c\! > \!J_\star[S]$ where $J_\star[3/2] \! \approx \! 3.3 J_1$
and $J_\star[1/2]\!\approx\! 0.66 J_1$.
Quantum Monte Carlo studies of this model would be valuable in firmly establishing the
value of $J_\star[S]$ as a function of $S$.
Fig.~\ref{fig:triplon} shows the triplon dispersion of the $S\!=\!3/2$ interlayer
VBS state at $J_c \!=\! 3.8 J_1$ (in units where $J_1\!=\!1$)  along high symmetry cuts
in the hexagonal Brillouin zone.
%; note the low energy
%triplon mode near the $\Gamma$-point.
For $J_c \! < \! J_\star[3/2]$,
or in the presence of a magnetic field which can close the spin gap in the VBS state for 
$J_c \!>\! J_\star[3/2]$, the low energy triplon
mode at the $\Gamma$-point condenses; its eigenvector is consistent with N\'eel order.

{\it Discussion.---} Motivated by recent experiments on Bi$_3$Mn$_4$O$_{12}$(NO$_3$), 
we have studied various honeycomb lattice spin models which support
quantum paramagnetic ground states that undergo
field-induced phase transitions to N\'eel order.
Detailed NMR
studies of isolated nonmagnetic impurities subsituted for Mn may help distinguish between 
these states. The interlayer VBS would have an impurity 
induced $S\!\!=\!\!3/2$ local
moment on the neighboring site in the adjacent layer, the AKLT state would 
nucleate three $S\!\!=\!\!1/2$ moments on neighboring sites
in the same plane, while spinless impurities in
spin gapped $Z_2$ fractionalized spin liquids \cite{fawang2010},
do not generically lead to local moments. 
Sharp dispersing triplet excitations expected in valence
bond solids discussed here could be looked for using single-crystal inelastic neutron scattering; 
by contrast, a spin liquid may not possess such sharp modes. 
Specific heat
experiments in a magnetic field could test for possible Bose-Einstein condensation
of triplet excitations as a route to N\'eel order.

Finally, dimer crystals with broken symmetry could also be candidate ground states
in Bi$_3$Mn$_4$O$_{12}$(NO$_3$) - in which case, disorder must be responsible for wiping 
out the thermal transition expected of such crystals. If BiMNO does support a valence bond 
solid ground state, disorder and Dzyaloshinskii-Moriya couplings (permitted by the bilayer 
structure) may be responsible for the observed nonzero low temperature susceptibility. This is 
an interesting direction for future research.

\acknowledgments

We thank
J. Alicea, M. Azuma, L. Balents, G. Baskaran, E. Berg, Y. B. Kim, M. Matsuda, and O. Starykh for discussions. 
This research was
supported by the Canadian NSERC (RG,YJ,AP), 
an Ontario Early Researcher Award (RG,AP), and 
U.S. NSF grants DMR-0906816 and DMR-0611562 (DNS).


\begin{thebibliography}{999}
\bibitem{reviews}
L. Balents, Nature {\bf 464}, 199 (2010);
R. Moessner and A.P. Ramirez, Phys. Today {\bf 59}, 24 (2006);
T. Giamarchi, Ch. R\"uegg, O. Tchernyshyov, Nat. Phys. {\bf 4}, 198 (2008).
\bibitem{BiMnO}
%O. Smirnova, M. Azuma, N. Kumada, Y. Kusano, M. Matsuda, Y. 
%Shimakawa, T. Takei, Y. Yonesaki, and N. Kinomura,
O. Smirnova, {\it et al}, J. Am. Chem. Soc., {\bf 131}, 8313 (2009);
%S. Okubo, F. Elmasry, W. Zhang, M.
%Fujisawa, T. Sakurai, H. Ohta, M. Azuma, O.
%A. Sumirnova, and N. Kumada,
S. Okubo, {\it et al},
J. Phys.: Conf. Ser. {\bf 200}, 022042 (2010).
\bibitem{fouet2001}
J. B. Fouet, P. Sindzingre, C. Lhuillier,
Eur. Phys. J. B {\bf 20}, 241 (2001).
\bibitem{mattsson1994}
A. Mattsson, P. Fr\"ojdh, and T. Einarson, \prb {\bf 49}, 3997 (1994).
\bibitem{takano2006}
K. Takano, \prb {\bf 74}, 140402 (2006).
\bibitem{assaad}
Z. Y. Meng, {\it et al}, Nature {\bf 464}, 847 (2010).
%T. C. Lang, S. Wessel, F. F. Assaad, A. Muramatsu, Nature {\bf 464}, 847 (2010).
\bibitem{mulder2010}
%A. Mulder, R. Ganesh, L. Capriotti, and A. Paramekanti, arXiv:1004.1119
%(unpublished).
A. Mulder, {\it et al}, Phys. Rev. B {\bf 81}, 214419 (2010).
\bibitem{kawamura2010}
%S. Okumura, H. Kawamura, T. Okubo, and Y. Motome, 
S. Okumura, {\it et al}, arXiv:1004.4441 (unpublished).
\bibitem{fawang2010}
F. Wang, \prb{\bf 82}, 024419 (2010);
Y.-M. Lu and Y. Ran, arXiv:1007.3266 (unpublished); B. K. Clark, D. A. Abanin, S. L. Sondhi,
arXiv:1010.3011 (unpublished).
\bibitem{jafari2010}
H. Mosadeq, F. Shahbazi, and S. A. Jafari, arXiv:1007.0127 (unpublished).
\bibitem{unpub}
M. Matsuda, {\it et al}, \prl {\bf 105}, 187201 (2010).
%M. Azuma, M. Tokunaga, Y. Shimakawa, and N. Kumada,
%(preprint).
\bibitem{AKLT1987}
%I. Affleck, T. Kennedy, E. H. Lieb, and H. Tasaki,
I. Affleck, {\it et al}, Phys. Rev. Lett. 59, 799 (1987); 
%I. Affleck, T. Kennedy, E. H. Lieb, and H. Tasaki,
I. Affleck, {\it et al}, Comm. Math. Phys. {\bf 115}, 477 (1988).
\bibitem{arovas1988}
D. P. Arovas, A. Auerbach, and F. D. M. Haldane, \prl {\bf 60}, 531 (1988).
\bibitem{KLT1988}
T. Kennedy, E. H. Lieb, and H. Tasaki, J. Stat. Phys. {\bf 53}, 383 (1988).
%\bibitem{akltoptical}
%J. Lavoie, {\it et al},
%Rainer Kaltenbaek, Bei Zeng, Stephen D. Bartlett, Kevin J. Resch
%Comments: 11 pages, 4 figures, 8 tables - added one reference
%Journal-ref: 
%Nature Physics {\bf 6}, 850 (2010).
\bibitem{alet2010}
A. F. Albuquerque, F. Alet, C. Sire, S. Capponi, Phys. Rev. B {\bf 81}, 064418 (2010).
\bibitem{brijesh}
B. Kumar, \prb {\bf 82}, 054404 (2010).
\bibitem{sachdevbhatt1990}
S. Sachdev and R. N. Bhatt, \prb {\bf 41}, 9323 (1990).
\bibitem{tcwei}
J. Cai, A. Miyake, W. D\"ur, and H. J. Briegel,
Phys. Rev. A {\bf 82}, 052309 (2010);
T.-C. Wei, I. Affleck, R. Raussendorf, arXiv:1009.2840 (unpublished).
\bibitem{disorderboson}
A. Paramekanti, N. Trivedi, and M. Randeria, \prb {\bf 57}, 11639 (1998).
\bibitem{Haldane}
F. D. M. Haldane, Phys. Lett. {\bf 93A}, 464 (1983); F. D. M. Haldane Phys. Rev. Lett. {\bf 50}, 1153 (1983).
\bibitem{young1989}
J. D. Reger, J. A. Riera, and A. P. Young, J. Phys. Cond. Matt.
{\bf 1}, 1855 (1989).
\bibitem{baskaran2009}
Z. Nourbakhsh, {\it et al}, J. Phys. Soc. Jpn. {\bf 78}, 054701 (2009).
\bibitem{square}
A. W. Sandvik and D. J. Scalapino
Phys. Rev. Lett. {\bf 72}, 2777 (1994); Y. Matsushita, M.  P. Gelfand, and C. Ishii, J. Phys. Soc. Jpn. {\bf 68}, 247 (1999).


%\bibitem{richter2004}
%J. Richter, J. Schulenberg, A. Honecker, and D. Schmalfuss, 
%J. Richter, {\it et al}, \prb{\bf 70}, 174454 (2004).
%\bibitem{bjyang2010}
%B.-J. Yang, A. Paramekanti, and Y. B. Kim,  Phys. Rev. B {\bf 81}, 134418 (2010).

\end{thebibliography}
\end{document}

\title{Supplementary Material for \\ ``Quantum paramagnets on the honeycomb lattice and field-induced
N\'eel order: Possible
application to Bi$_3$Mn$_4$O$_{12}$(NO$_3$)"}
\author{R. Ganesh}
\affiliation{Department of Physics, University of Toronto, Toronto, Ontario M5S 1A7, Canada}
\author{D. N. Sheng}
\affiliation{Department of Physics and Astronomy, California State University, Northridge, California 91330, USA}
\author{Y. J. Kim}
\affiliation{Department of Physics, University of Toronto, Toronto, Ontario M5S 1A7, Canada}
\author{A. Paramekanti}
\affiliation{Department of Physics, University of Toronto, Toronto, Ontario M5S 1A7, Canada}
\affiliation{Canadian Institute for Advanced Research, Toronto, Ontario, M5G 1Z8, Canada}
\maketitle

\section{Spin wave fluctuations around canted N\'eel state}

The Hamiltonian, with a magnetic field and second neighbour exchange, is
\bea
\label{Eq:Hnnexch}
H = J_1 \sum_{\la i j\ra} \bS_i \cdot \bS_j + J_2  \sum_{\la\la i j \ra\ra} \bS_i \cdot \bS_j  - B
\sum_i S_{i}^{z}.
\eea
As discussed in the main body, when $J_2 < J_1/6$ and $B < 6J_1 S$, N\'eel ordering is in the plane perpendicular to the magnetic field, but
the spins also uniformly cant in the direction of applied field, to maximally gain Zeeman energy.
The classical spin state can be characterized by $\bS_\br \!=\! S (\pm \cos\chi,0,\sin\chi)$ on the two sublattices.
We now define new spin operators, denoted by ${\mathbf T}_{i,\alpha}$, via a sublattice-dependent local spin rotation 
\bea
\left(\begin{array}{c}
       T_{i,\alpha}^x \\ T_{i,\alpha}^y \\ T_{i,\alpha}^z
      \end{array} \right)\!\! =\!\!
\left(\begin{array}{ccc}
       \sin\chi & 0 & (-)^{\alpha+1}\cos\chi \\
	0 & 1 & 0 \\
	(-)^\alpha \cos\chi & 0 & \sin\chi \end{array}\right)\!\!
\left(\begin{array}{c}
       S_{i,\alpha}^x \\ S_{i,\alpha}^y \\ S_{i,\alpha}^z
      \end{array} \right),
\eea
where $\alpha=1,2$, is a sublattice index and $i$ sums over each unit cell.

The ground state has all spins pointing towards the new local-$S^z$ axis. To study spin wave fluctuations, we rewrite the 
$T$ operators in terms of Holstein-Primakoff bosons as follows:
\bea
\nn T_{i,\alpha}^{z}\!\! &=& \! S-b_{i,\alpha}^\dg b_{i,\alpha}, \\
\nn T_{i,\alpha}^{x}\!\! &=& \!\sqrt{\frac{S}{2}}(b_{i,\alpha}+b_{i,\alpha}^\dg), \\
\nn T_{i,\alpha}^{y}\!\! &=& \!\frac{1}{i}\sqrt{\frac{S}{2}}(b_{i,\alpha}-b_{i,\alpha}^\dg). 
\eea

The Hamiltonian can now be rewritten as $H\approx E_{Cl}+ H_{qu}$. The classical energy $E_{Cl}$ is proportional to $S^2$, and the leading order quantum correction, $H_{qu}$, is of order $S$. We get the value of the canting angle $\chi$ by demanding that terms of order $S^{3/2}$, which are linear in the boson operators, should vanish, which yields
\bea
\sin\!\chi = \frac{B}{6J_1 S}.
\eea
The classical energy is given by 
\bea
\frac{E_{Cl} }{N S^2}=-\frac{3}{2}J_1 \cos2\chi + \frac{3}{2}J_2 - \frac{B}{S} \sin\!\chi.
\eea
where $N$ is the number of sites in the honeycomb lattice. We take the magnetic field $B$ to be of order $S$, so that the Zeeman 
term $-BS_{i}^{z}$ is treated on the same level as the exchange terms $J_{ij} \bS_{i}\!\cdot\!\bS_{j}$.
The leading quantum correction
\bea
\nn \frac{H_{qu}}{S N} &=& -\frac{3}{2}J_1\cos2\chi+3J_2 -\frac{B}{2S}\sin\chi    \\
&+&\sum_{\bk>0} \psi_{\bk}^\dg H_{\bk} \psi_{\bk},
\eea
where
\bea
\psi_{\bk} = \left(\begin{array}{c}
       b_{\bk,1} \\ b_{\bk,2} \\ b_{-\bk,1}^\dg \\ b_{-\bk,2}^\dg
      \end{array}\right);
%\left(\begin{array}{c}
%       a_{\bk}^\dg \\ b_{\bk}^\dg \\ a_{-\bk} \\ b_{-\bk}
%      \end{array}\right)^{T}
H_{\bk}=\left(\begin{array}{cccc}
I_{\bk} & F_{\bk} & 0 & G_{\bk}\\
F_{\bk}^* & I_{\bk} & G_{\bk}^* & 0 \\
0 & G_{\bk} & I_{\bk} & F_{\bk} \\
G_{\bk}^* & 0 & F_{\bk}^* & I_{\bk}
      \end{array}\right)
\eea
with
\bea
\nn I_{\bk} &=& 3J_1 \cos2\chi - 6J_2,  \\
\nn &+& \!\! 2J_2\{ \cos k_a + \cos k_b + \cos(k_a+k_b)\}\! + \!\frac{B}{S}\sin\chi, \\
\nn F_{\bk} \! &=& \!\!J_1 \gamma_{\bk} \sin^2 \chi  \equiv \vert F_{\bk} \vert e^{i\eta_{\bk}},\\
\nn G_{\bk}\!&=&\!\! -J_1 \gamma_{\bk} \cos^2\chi, 
\eea
where $\gamma_\bk$ is as defined in the main body. This Hamiltonian can be diagonalized by a bosonic Bogoliubov transformation. The eigenvalues are given by 
\bea
\Omega_{\bk}^{\pm} = \sqrt{(I_{\bk}\pm \vert F_{\bk} \vert)^2 - \vert G_{\bk}\vert^2}.
\eea
The transformation matrix is given by 
\bea
\nn P=\left(\begin{array}{cc}
U_{2\times2} & 0 \\
0 & U_{2\times2}
\end{array}\right)
\left(\begin{array}{cc}
C_{2\times2} & S_{2\times2} \\
S_{2\times2} & C_{2\times2}
\end{array}\right),
\eea
where 
\bea
U_{2\times2}=\frac{1}{\sqrt{2}}\left(\begin{array}{cc}
-e^{i\eta_{\bk}} & e^{i\eta_{\bk}}\\
1 & 1 
%-\frac{F_{\bk}}{\sqrt{2}\vert F_{\bk}\vert} & \frac{F_{\bk}}{\sqrt{2}\vert F_{\bk}\vert} \\
%\frac{1}{\sqrt{2}} & \frac{1}{\sqrt{2}} 
\end{array}\right);
\eea
\bea
C_{2\times2}\!\!=\!\!\left(\begin{array}{cc}
\cosh\theta & 0 \\
0 & \cosh\phi
\end{array}\right)\!;\!
S_{2\times2}\!\!=\!\!\left(\begin{array}{cc}
\sinh\theta & 0 \\
0 & \sinh\phi
\end{array}\right),
\eea
where the angles $\theta$ and $\phi$ are given by
\bea
\nn \tanh2\theta = \frac{\vert G_{\bk} \vert}{I_{\bk}-\vert F_{\bk}\vert}, \\
\nn \tanh2\phi = \frac{-\vert G_{\bk}\vert}{I_{\bk}+\vert F_{\bk}\vert}. \\
\eea
The matrix P preserves the commutation relations of the bosonic operators and diagonalizes the Hamiltonian, giving $P^\dg H P = {\rm Diag}\{\Omega_{\bk}^{-},\Omega_{\bk}^{+},\Omega_{\bk}^{-},\Omega_{\bk}^{+} \}$.

The expectation value of spin moments can be calculated in terms of this new basis. For example, the in-plane component of the spin is given by
\bea
\nn\frac{1}{N}\sum_i \la S_{i,\alpha=1}^{x}\ra\!\! =\!\! (S+1/2)\cos\chi\!-\frac{\cos\chi}{2N}\times \\
\sum_{\bk>0}\!\left[
 \cos\!2\theta \{1\!\!+\!2n_B (\Omega_{\bk,-}\!)\!\}\!
\!+\!\cos\!2\phi \{1\!\!+\!2n_B\! (\Omega_{\bk,+})\}\right]\!\!,
\eea
where $n_B(.)$ denotes the Bose distribution function. For $T\neq 0$, a small coupling along the third dimension is necessary to allow for a stable magnetically ordered state. We take this into account by imposing an infrared cutoff $\Lambda$ which is of the order the interplane couling; modes with energy greater than $\Lambda$ look like 2d spin waves. 

\begin{figure}[tb]
\includegraphics[width=3.3in]{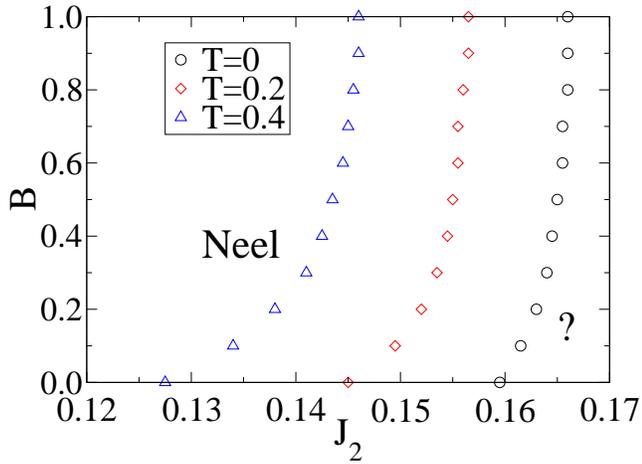}
\caption{Melting of N\'eel order for $S\!\!=\!3/2$ in the 
$B$-$J_2$ plane for $T=0, 0.2, 0.4$ (in units of $J_1$). To the left of the curve, there is stable canted N\'eel order. To the right, fluctuations melt the in-plane N\'eel order. 
}
\label{fig:finiteTmelting}
\vskip -0.1in
\end{figure}

As $J_2$ is increased, fluctuations around the N\'eel state increase. We expect the N\'eel state to melt when fluctuations become comparable to the magnitude of the ordered moment. As discussed in the main body, we choose our melting criterion to $\sqrt{\la S_x^2\ra \!-\! \la S_x\ra^2} \!\! = \!\! 3\la S_x\ra$, which gives a critical $J_2/J_1$ in agreement with recent variational Monte Carlo results for $S=1/2$.
 The resulting melting curve for $S=3/2$, at zero temperature and at small temperatures, is shown in Fig.\ref{fig:finiteTmelting}.

%\begin{thebibliography}{999}
%\bibitem{VariationalMC}
%B. K. Clark, D. A. Abanin, S. L. Sondhi, arXiv:1010.3011 (unpublished).
%\end{thebibliography}